# Noise-Adaptive Intelligent Programmable Meta-Imager

Chenqi Qian[1] and Philipp del Hougne[1*]

[1] Univ Rennes, CNRS, IETR - UMR 6164, F-35000 Rennes, France

* Correspondence to philipp.del-hougne@univ-rennes1.fr.

## Abstract

We present an intelligent programmable computational meta-imager that tailors its sequence of coherent scene illuminations not only to a specific information-extraction task (e.g., object recognition) but also adapts to different types and levels of noise. We systematically study how the learned illumination patterns depend on the noise, and we discover that trends in intensity and overlap of the learned illumination patterns can be understood intuitively. We conduct our analysis based on an analytical coupled-dipole forward model of a microwave dynamic metasurface antenna (DMA); we formulate a differentiable end-to-end information-flow pipeline comprising the programmable physical measurement process including noise as well as the subsequent digital processing layers. This pipeline allows us to jointly inverse-design the programmable physical weights (DMA configurations that determine the coherent scene illuminations) and the trainable digital weights. Our noise-adaptive intelligent meta-imager outperforms the conventional use of pseudo-random illumination patterns most clearly under conditions that make the extraction of sufficient task-relevant information challenging: latency constraints (limiting the number of allowed measurements) and strong noise. Programmable microwave meta-imagers in indoor surveillance and earth observation will be confronted with these conditions.

## Introduction

Any measurement process is inevitably corrupted by noise of some type and level. We hypothesize that the optimal coherent illumination patterns to be used by an intelligent programmable meta-imager to efficiently extract task-specific information from a scene will depend on the type and level of noise (*1*). We consider multi-shot programmable computational-imaging systems based on a single transmitter and a single detector; such systems are of particular relevance to the microwave domain where transceivers are costly and programmable metasurface apertures can synthesize (capture) coherent wavefronts from (with) a single radiofrequency chain. Envisioned deployment scenarios of programmable microwave meta-imagers include indoor surveillance and earth observation, where stringent latency constraints (limiting the number of allowed measurements) and significant amounts of noise combined with weak signals require that the utilized sequence of scene illuminations optimally extracts task-relevant information.

To systematically evaluate our hypothesis, we consider a prototypical object-recognition problem in which one microwave dynamic metasurface antenna (DMA) radiates a sequence of coherent wavefronts to the scene, and a second DMA coherently captures the reflected waves. For a given number of allowed measurements, we learn a task-specific and noise-specific sequence of optimized DMA configurations and benchmark its performance against conventional compressed sensing with random configurations. Our learned-sensing approach is advantageous whenever the amount of information that can be extracted from the scene is limited through latency constraints and/or noise such that it becomes crucial to maximize the amount of task-relevant information in the measured data. Moreover, we investigate how overlap and intensities of the learned illumination patterns depend on the noise type and level. We discover that these trends are intuitively understandable despite the black-box nature of the learned-sensing approach and the complexity of both metamaterial hardware and recognition task.

**Compressive vs. Intelligent Meta-Imagers.** Computational imagers multiplex scene information across a "coded aperture" onto a detector such that there is no one-to-one mapping between scene voxels and measured data. The coded aperture can be part of the transmit and/or receive hardware. Compressive imagers, a subset of computational imagers, take advantage of the inherent sparsity of most natural scenes to linearly embed the scene in a lower-dimensional space. Random masks were shown to enable such isometric embeddings which are distance-preserving transformation; in other words, multiplexing the scene information across random masks compresses the scene information without distorting it. The ability to recover a scene from highly incomplete measurements sparked considerable interest, and much effort in the metamaterials community went into conceiving metamaterial hardware that can emulate random masks – see Sec. II in Ref. (*1*). Importantly for the current work, Ref. (*2*) introduced programmable meta-imager hardware that can synthesize or capture different pseudo-random coherent wavefronts at a single frequency by randomly reconfiguring *in situ* the scattering properties of programmable meta-atoms in the transceiver hardware.

Two broad classes of programmable meta-imager hardware exist. A first class of programmable meta-imagers consists of structures that guide or trap waves; these structures are patterned with individually addressable meta-atoms that leak out (capture) wave energy toward (from) the scene via those meta-atoms that are configured to be resonant. The DMAs considered in this work are 2D parallel-plate waveguides with sub-wavelength 1-bit programmable meta-atoms patterned into one of the conducting surfaces (*3*). A second class of programmable meta-imagers consists of programmable metasurface reflect- or transmit-arrays, nowadays also coined reconfigurable intelligent surfaces, that sculpt waves from a feed antenna in order to flexibly illuminate a scene (*4*). Programmable meta-imagers open the door for various kinds of optimizations of the used configurations to illuminate the scene, going beyond the use of random configurations from Refs. (*2*, *4*). For example, the configurations can be optimized to minimize the overlap of subsequent scene illuminations and thereby avoid the acquisition of redundant information (*5*, *6*). It is also possible to use a specific, as opposed to arbitrary, orthogonal basis of scene illuminations in which a typical scene can be expressed as linear superposition of a small number of "principle" patterns (*7*, *8*).

The use of meta-imager configurations yielding pseudo-random, orthogonal or principle-component-based scene illuminations isometrically embeds *all* scene information in a lower-dimensional space and is hence by definition agnostic to the task and the noise. However, many deployment scenarios involve latency constraints and noise, both of which limit the amount of information that can be extracted from the scene. In such situations, it is desirable to instead use a purposefully non-isometric embedding that *discriminates between task-relevant and task-irrelevant* information as much as possible. This learned-sensing approach uses the same programmable meta-imager hardware but judiciously tailors the utilized configurations such that the measurement process becomes simultaneously a *task-specific* "over-the-air" analog processing step. The selection of task-relevant information during the measurement, as opposed to an indiscriminate acquisition of all information, distinguishes intelligent meta-imagers from compressive meta-imagers (*1*). The concept of intelligent meta-imaging was introduced in Ref. (*9*) where an end-to-end pipeline including both the physical measurement process, parametrized by the meta-atom configurations, and the digital processing layer, parametrized by digital weights, was optimized with respect to the performance on a specific task. The resulting task-aware meta-imager configurations yielded substantial accuracy improvements under latency constraints, a result that was subsequently confirmed experimentally in Ref. (*10*).

However, while Refs. (*9, 10*) learned task-specific configurations for their programmable meta-imagers, they were limited to scenarios without noise or with negligible noise, respectively. Yet, according to our aforementioned hypothesis, noise may profoundly impact the optimal meta-imager configurations because, besides latency constraints, noise also limits the amount of information that can be extracted from the scene. As stated above, many envisioned microwave meta-imaging applications will inevitably be confronted with strong noise. Signal averaging to reduce the noise may not be optimal due to the inevitable latency penalty associated with repeated measurements. Moreover, in the remaining scenarios, one may desire to trade off the emitted signal power against performance in order to improve metrics like energy consumption, radiation exposure and spectrum allotment. Hence, understanding how the performance of intelligent meta-imagers depends not only on latency constraints but also on the noise is important.

**Noise in Task-Specific Sensing.** A related strand of research in the optical domain already routinely includes noise in similar end-to-end task-specific optimizations of physical hardware and digital processing layers (*11–17*). However, typically only a single low noise level is (sometimes somewhat arbitrarily) chosen and optimized for in these works, such that there is no systematic investigation of the effect of noise, especially not of strong noise, on the optimized illumination patterns. Moreover, these optical learned-sensing schemes differ conceptually from the present work in that they operate with a *single-shot detector-array* measurement. By construction, these schemes hence do not face a latency-induced limitation on the amount of information that can be extracted from the scene, unlike the typical *multi-shot single-detector* programmable meta-imagers considered in the present work that must limit their number of measurements to avoid prohibitive latencies. This difference originates from the fact that the optical regime is confronted with different hardware constraints: while it is costly to measure phase, detector arrays like CCD cameras are available at low or moderate cost. Only a single illumination pattern is hence optimized in typical optical learned sensing, precluding the investigation of intriguing aspects such as the overlap of subsequent learned illumination patterns in multi-shot schemes that are discussed in the present work. Moreover, this single illumination pattern is often implemented with a static optimized optical element (e.g., a metasurface) as opposed to an in-situ reconfigurable measurement setup as considered in this work. Departing from the single-shot approach, Refs. (*18, 19*) from the optical domain optimized exactly two illumination patterns for a specific task and one chosen noise level; again, no systematic analysis of the role of noise on the learned illumination patterns was presented and strong-noise regimes were not considered.

The complexity of typical information-extraction problems implies that there is, in general, no guarantee to converge toward the globally optimal set of scene illuminations. However, heuristic evidence suggests that different realizations converge to distinct local optima of roughly equal quality (*9*). Nonetheless, in a few special cases, provably optimal "maximum information states" can be identified analytically. If the task is to distinguish between exactly two specific scene configurations without any realization-to-realization variation using a single-shot coherent illumination, and assuming that the difference $\Delta T$ between the multi-channel input-output relation of the system can be measured *without any noise* for these two configurations, then it follows from basic linear algebra that the first eigenvector of $(\Delta T)^\dagger (\Delta T)$ is the optimal illumination pattern (*20*). A similar technique can be applied to the precise estimation of a small perturbation of a single parameter in the vicinity of a known value (*21*). However, we are generally interested in sensing non-perturbative changes with a certain degree of realization-dependence and more than two possible outcomes, and without reliance on noise-free characterization measurements. Therefore, the identification of task-specific illumination patterns generally involves an optimization problem.

**Our Contributions.** In this paper, we systematically explore how the combination of latency constraints and noise impacts intelligent multi-shot programmable meta-imagers. Considering microwave DMA hardware for a prototypical object-recognition problem, we benchmark the performance of a noise-adaptive intelligent programmable meta-imager against the conventional use of random configurations. We consider the entire range of conceivable noise levels for two distinct noise types: signal-*independent* additive Gaussian noise and signal-*dependent* additive Gaussian noise. We analyze how the average overlap and the average intensity of the learned scene illuminations vary with the number of allowed measurements, the noise level, and the noise type. Thereby, we discover intuitively understandable trends.

This paper is organized as follows. First, we describe the considered system model. Second, we formulate the end-to-end information flow through our wave-based information-extraction problem. Third, we discuss the end-to-end task-specific optimization. Fourth, we present our analysis of performances and illumination-pattern sequences for the case of signal-*independent* additive noise. Fifth, we perform the same analysis for the case of signal-*dependent* additive noise. Sixth, we study the performance outside the trained noise regime. Finally, we conclude with a discussion and summary.

## Results

**System Model.** We consider the microwave computational programmable meta-imager system introduced in Ref. (*9*). As illustrated in Figure 1, a scene in free space[1] is exposed to a sequence of *M* coherent illuminations generated by a transmitting (TX) DMA, and the reflected waves are coherently captured by a second receiving (RX) DMA. The DMAs are parallel-plate waveguide structures connected to a single-mode coaxial cable and patterned with programmable meta-atoms on one of their surfaces (*3, 9*). Each meta-atom is individually 1-bit programmable and approximated as a discrete dipole; in its ON (OFF) state, a programmable meta-atom is (is not) resonant and hence has a finite (zero) dipole moment such that it couples (does not couple) waveguide modes to modes propagating in free space. The dipole moment of a resonant meta-atom depends on the incident field and hence on (i) its location on the waveguide surface, and (ii) the configuration of the remaining meta-atoms due to multiple scattering. The *i*th measurement yields a single complex-valued scalar $m_i$ that is corrupted by additive noise $n_i$.[2] Under the first Born approximation,

$$m_i = \int_{\text{scene}} E_i^{\text{TX}}(\mathbf{r}) E_i^{\text{RX}}(\mathbf{r}) \sigma(\mathbf{r}) d\mathbf{r} = \int_{\text{scene}} \mathcal{I}_i(\mathbf{r}) \sigma(\mathbf{r}) d\mathbf{r}, \qquad (1)$$

where $\mathbf{r}$ denotes a coordinate in 3D space, $E_i^{\text{TX}}(\mathbf{r})$ and $E_i^{\text{RX}}(\mathbf{r})$ are the electric field patterns of the TX DMA and the RX DMA in the plane of the scene, respectively, and $\sigma(\mathbf{r})$ is the scene reflectivity (*23*). Moreover, we define the *i*th coherent illumination pattern $\mathcal{I}_i(\mathbf{r}) = E_i^{\text{TX}}(\mathbf{r}) E_i^{\text{RX}}(\mathbf{r})$. I and Q components of the *M*-element complex-valued vector of measured data, $[m_1 + n_1, m_2 + n_2, \dots, m_M + n_M]$, are stacked and normalized to zero mean and unity standard deviation before being fed into a fully-connected artificial neural network.

---

[1] Computational meta-imagers are to date mainly studied for operation in free space where the Green's function between metasurface aperture and scene is known analytically. Interestingly, first investigations of computational meta-imaging in complex scattering environments indicate that the reverberation in such conditions can strongly improve the achievable precision because it introduces a generalized interferometric sensitivity (*22*).

[2] In principle, the noise can also be non-additive (e.g., in Poisson-distributed processes). In the most general formulation, the measured complex-valued scalar is $f(m_i)$, where $f(\cdot)$ denotes the arbitrarily complex noise function acting on $m_i$.

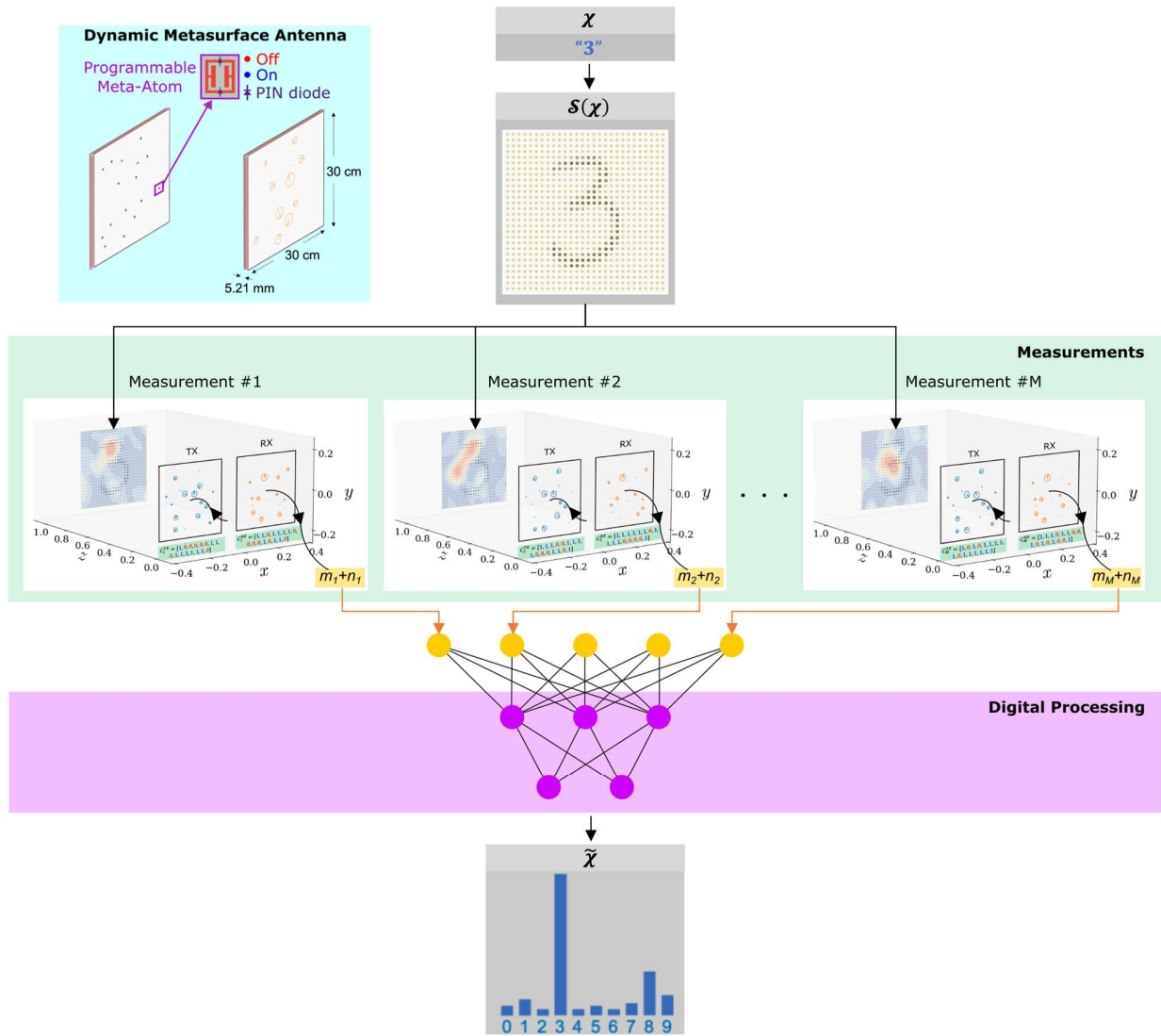

**Figure 1: Schematic Overview.** An information of interest $\chi$, "3", is physically encoded in a scene reflectivity of a corresponding handwritten metallic digit. The scene is probed with *M* measurements. For each measurement, a TX DMA radiates a coherent field toward the scene, and the reflection is coherently captured by a RX DMA, yielding a single complex-valued scalar $m_i$ that is corrupted by noise $n_i$. The coherently radiated and captured fields of the TX and RX DMAs in the *i*th measurement are determined by the configuration of the DMAs' meta-atoms, $C_i^{\mathrm{TX}}$ and $C_i^{\mathrm{RX}}$, respectively. The wave energy injected into the TX DMA is always the same. The measured data from the *M* measurements is injected into a fully-connected digital-processing neural network in order to output an estimate $\tilde{\chi}$ of the sought-after information of interest. The overall procedure is hence parametrized by physical weights (the meta-atom configurations) and digital weights (the digital ANN weights). The inset shows the considered DMA hardware: a 2D parallel-plate waveguide with 16 1-bit programmable meta-atoms patterned into the surface facing the scene. Depending on the bias voltage applied across the PIN diode, a given meta-atom is resonant (blue) or not (red), as shown for an example configuration on the left. The corresponding dipole moments of the meta-atoms are shown on the right as phasors.

In the present work, we describe the entire physical measurement process with an analytical coupled-dipole forward model that approximates the subwavelength meta-atoms as dipoles. This involves determining the dipole moments of each meta-atom for each configuration (*3*), as well as the use of Eq. (1) to propagate the field from the TX DMA to the scene, and back from the scene to the RX DMA (*23*). We assume that the amount of wave energy injected into the TX DMA is always the same throughout this work. For the sake of brevity, we do not repeat the mathematical details of our analytical forward model here; instead, we refer the reader to Ref. (*9*) where they are provided and explained in depth. Note that our results presented below rely on having a differentiable forward model, but there is no requirement for this model to be analytical; instead, one could, for instance, also learn a forward model (*10*). For future experimental implementations, a physics-based learned forward model of the physical measurement process could be an enticing option to exploit partial physics knowledge in combination with fine tuning learned from calibration data of the experimental system (*24*).

**Information Flow.** To implement a task-specific end-to-end optimization of both physical and digital layers, a prerequisite is the formulation of the pipeline through which information flows in the considered problem. The essential elements of this pipeline are the same for all wave-based information-extraction problems, including imaging, sensing, localization, and object recognition. The only significant difference lies in the task-specific cost function that is to be optimized for good performance. Two broad classes of tasks can be distinguished: classification-type tasks and regression-type tasks. Classification-type tasks include, for instance, object recognition on which we focus for concreteness in the following. For classification tasks, the cost function determines the precision with which a given scene is correctly determined to belong to one out of $P$ pre-defined classes. Regression-type tasks include, for instance, the estimation of a continuous parameter (e.g., the location of an object) as well as image reconstruction. For regression tasks, the cost function quantifies the difference between the true and estimated variable(s), such as an object location or the pixel values in an image.

The starting point of the information flow is the sought-after latent information of interest $\chi$. The latter exists in some abstract latent information space. In our prototypical example, we seek to recognize metallic digits in the scene. The latent information of interest is hence the fact that the digit belongs to one of the $P = 10$ possible classes. As illustrated in Figure 2, the latent information of interest could hence, for instance, be "3". Some scene function $\mathcal{S}$ maps the latent information space to the experimental reality by encoding $\chi$ in the scene reflectivity $\sigma(\mathbf{r})$. This scene function $S$ usually describes some natural process, in our case handwriting, that can have a certain realization-to-realization variation (i.e., different people write digits to some extent differently) and is, in general, *not* parametrized by any controllable parameters that could be optimized. The scene reflectivity $\sigma(\mathbf{r})$ is hence the physically encoded information of interest, $S(\chi)$.

The scene is now probed with $M$ coherent illumination patterns $\{\mathcal{I}_i(C_i)\}$ which are generated with $M$ configurations $\{C_i\}$ of the DMAs, as detailed in the previous section.[3] Given the single-detector nature of the considered measurement process, each measurement yields a single complex-valued scalar measurement that depends on both the scene and the utilized configuration of the TX DMA and the RX DMA: $m_i(S(\chi), C_i)$. In addition, the measurement is corrupted by additive noise $n_i$, as mentioned previously. The measurement process is hence parametrized by trainable physical weights, the $M$ DMA configurations $\{C_i\}$, which are highlighted in green font in Figure 2 and can be optimized. Overall, we summarize the mapping of the scene $S(\chi)$ to the measured data vector through the measurement function $\mathcal{M}$.

Finally, a digital-processing function $\mathcal{D}$ acts on the measurement data to obtain an estimate $\tilde{\chi}$ of the information of interest. This digital processing step is parametrized by a set of trainable digital parameters $\{\phi_i\}$, for instance, the weights of a fully connected artificial neural network (ANN) in our present work, which are highlighted in pink font in Figure 2 and can be optimized.

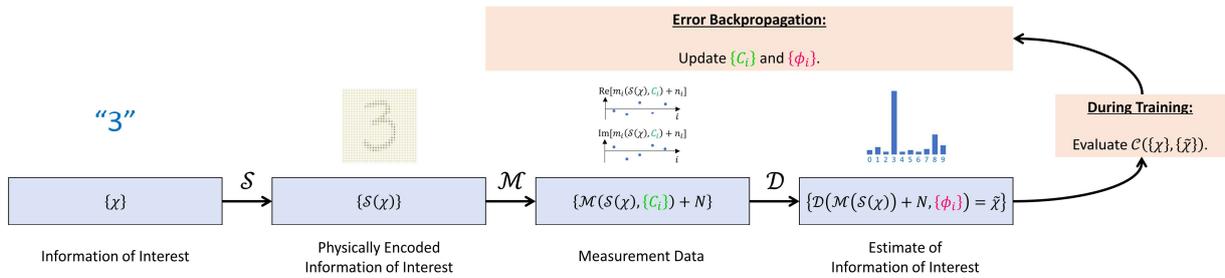

**Figure 2: Information-Flow Pipeline for Task-Specific and Noise-Adaptive Coherent Wave-Based Information Extraction.** A batch $\{\chi\}$ of instances of the latent information of interest $\chi$, an example being "3" for our handwritten digit recognition task, is mapped via the scene function $S$ to the real world; in our case, the scene reflectivity physically encodes the information of interest. Next, a measurement function $\mathcal{M}$, parametrized by a set of physical parameters $\{C_i\}$ (in our case the sequence of DMA configurations; green), maps the scene reflectivity to a complex-valued vector of measured noise-corrupted data. Finally, a digital-processing function $\mathcal{D}$, parametrized by a set of digital parameters $\{\phi_i\}$ (in our case the ANN weights; pink), attempts to extract an estimate $\tilde{\chi}$ from the measured data. The cost function evaluates how close the

---

[3] For the sake of compactness, we denote with $C_i = \{C_i^{\mathrm{TX}}, C_i^{\mathrm{RX}}\}$ jointly the $i$th configuration of the TX DMA, $C_i^{\mathrm{TX}}$, and the $i$th configuration of the RX DMA, $C_i^{\mathrm{RX}}$.

corresponding batches of $\{\chi\}$ and $\{\tilde{\chi}\}$ are, in order to subsequently backpropagate the error for a *joint* optimization of the physical and digital weights, $\{C_i\}$ and $\{\phi_i\}$.

**End-to-End Task-Specific Optimization.** Having established our system model and having formulated the information-flow pipeline, we now tackle the task-specific end-to-end joint optimization of the trainable physical parameters (DMA configurations) and trainable digital parameters (ANN weights). It is this *joint* optimization that endows the measurement process with task awareness, such that it can discriminate between task-relevant and task-irrelevant information over the air in the analog domain. These features distinguish intelligent meta-imagers from compressive meta-imagers (*1*). Compressive meta-imagers may optimize the $M$ DMA configurations $\{C_i\}$, but without taking the task or the noise into account.

In principle, any optimization algorithm can be used to jointly optimize physical and digital weights for a specific task and noise. Yet, given the differentiable forward model and the ANN on the digital-processing layer, the most convenient approach is to interpret the entire information-flow pipeline as one hybrid analog-digital neural network such that error backpropagation algorithms from well-established libraries can be used to update the physical and digital weights in order to minimize the task-specific cost function $\mathcal{C}(\{\chi\}, \{\tilde{\chi}\})$, where $\{\cdot\}$ denotes a batch of training data. The noise which corrupts the measurements in the information-flow pipeline is obviously of statistical nature and hence realization-dependent. In other words, during every training iteration's forward pass through the pipeline a new noise realization from a chosen noise distribution is used, such that the algorithm can adapt to the statistical properties of the noise distribution rather than being specific to one noise realization.

One complication of typical programmable meta-imager hardware is the 1-bit or few-bit (as opposed to continuous) programmability of the meta-atoms that may appear incompatible with error backpropagation. To overcome this challenge, we follow Refs. (*9, 11*) and use a "temperature parameter" that gradually drives the values of the trainable physical parameters from a continuous distribution to the desired discrete distribution. Algorithmic details are available in Ref. (*9*). We use the MNIST with 70,000 examples of handwritten digits for training (51,000), validation (9,000) and testing (10,000).

**Signal-Independent Additive Noise.** We begin by considering the most common model for noise in the kind of microwave measurements we are concerned with: signal-independent additive Gaussian noise. Specifically, we consider independent and identically distributed zero-mean Gaussian noise with standard deviation $\rho$ on the I and Q components of the complex-valued measured data. We define the units of $\rho$ such that $\rho = 1$ corresponds to an SNR of 0 dB if random DMA configurations are used, where the SNR is defined as the ratio of the signal variance to the noise variance. It is important to note that the same value of $\rho$ can correspond to different values of SNR if non-random DMA configurations are used: although always the same amount of wave energy is injected into the TX DMA, the received signal level could be significantly higher or lower depending on (i) how many meta-atoms are configured to be resonant ("ON"), and (ii) whether they create constructive or destructive interferences in the illumination patterns. In the most extreme case with all TX DMA and/or all RX DMA elements configured to be non-resonant, zero signal would be captured, and, irrespective of the value of $\rho$, the SNR would tend to negative infinity. For these reasons, we perform our analysis in terms of $\rho$ as opposed to SNR. Moreover, we note that considering non-zero-mean Gaussian distributions would yield the same results in our system model due to (i) the above-mentioned normalization of the measured data before it is fed into the digital layers and (ii) the fact that we assume lossless analog-to-digital conversion (*25*), which is effectively available, for instance, with commercially available vector network analyzers in the microwave domain.

We expect the use of end-to-end optimized task-specific DMA configurations to outperform the conventional use of random configurations in conditions under which only a limited amount of information can be extracted from the scene: when the number of allowed measurements $M$ is limited due to latency constraints and/or when the noise is strong. Therefore, we explore the performance of these two approaches (learned illuminations vs. random illuminations) as a function of $M$ and $\rho$ in the following. To begin, we consider the achieved accuracy as a function of $M$ for the three representative noise levels of $\rho = 0.1, 1$ and $10$. The corresponding results are plotted in the top row of Figure 3 and show the average and standard deviation over 10 separate optimizations for each combination of $M$ and $\rho$. To avoid that the results are specific to a DMA layout, we choose different random locations of the 16 meta-atoms on the TX and RX DMA apertures for each of the ten optimizations.

Let us begin by examining the extreme cases. For a low noise level ($\rho = 0.1$) and many measurements ($M > 10$), both approaches achieve close to unity accuracy because sufficient task-relevant information is included in so many measurements even if they are not task-specific. For a high noise level and few measurements, the results of both techniques approach the random-guess baseline of 10 % accuracy (the $P = 10$ classes are equiprobable). Using few and very noisy measurements does not allow the system to extract sufficient task-relevant information to achieve good performance, although the use of learned patterns remains superior to the use of random patterns, as expected. The most interesting and application-relevant regime is between these two extreme cases. We note that the accuracy with the learned illuminations is always superior to the random-illumination benchmark. For $\rho = 10$, the learned patterns outperform the random patterns most clearly for the largest considered value of $M = 150$ (46 % vs. 18 %). For $\rho = 1$, the largest absolute performance gain from using learned patterns is seen at $M = 8$ (72 % vs. 40 %). For $\rho = 0.1$, the largest absolute performance gain is at $M = 2$ (70 % vs. 50 %) and $M = 3$ (83 % vs. 63 %). The performance gap peak hence occurs at lower values of $M$ for lower values of $\rho$. Naturally, the advantage of learned patterns over random patterns occurs in a regime in which the total amount of information that can be extracted from the scene is limited – but not too limited such that even the learned patterns struggle to capture sufficient task-relevant information.

We now aim to gain insights into *how* the end-to-end task-specific optimization achieves these remarkable performance improvements over the conventional random patterns. Given the complexity of the DMA hardware as well as the recognition task, we do not hope to comprehend "microscopically" every meta-atom's specific learned configuration; instead, we now analyze "macroscopically" each sequence of $M$ illumination patterns, $\{\mathcal{I}_i\}$, in terms of the illumination patterns' mutual overlaps and their intensities. Specifically, we evaluate the mean illumination pattern overlap within a given sequence,

$$\mathcal{O} = \left\langle \left| \frac{\int_{\text{scene}} \mathcal{I}_i(\mathbf{r})\mathcal{I}_j(\mathbf{r})\mathrm{d}\mathbf{r}}{\sqrt{\int_{\text{scene}} \mathcal{I}_i(\mathbf{r})\mathcal{I}_i(\mathbf{r})\mathrm{d}\mathbf{r} \int_{\text{scene}} \mathcal{I}_j(\mathbf{r})\mathcal{I}_j(\mathbf{r})\mathrm{d}\mathbf{r}}} \right| \right\rangle_{i \neq j}, \quad (2)$$

and the corresponding mean illumination pattern intensity,

$$I = \left\langle \int_{\text{scene}} |\mathcal{I}_i(\mathbf{r})|^2 \mathrm{d}\mathbf{r} \right\rangle_i. \quad (3)$$

In the second row of Figure 3, we plot the average over the ten realizations of the mean illumination pattern overlap $\mathcal{O}$. If random illumination patterns are used, the average overlap is by construction independent of $M$ and $\rho$ (since neither of these influences the choice of illumination patterns) at a value of 29 %. For the low-noise regime ($\rho = 0.1$), the average overlap of the learned patterns is identical to that of the random patterns and also independent of $M$. This is consistent with the findings in Ref. (*9*) for noiseless operation. Remarkably, concerns about the acquisition of redundant information in subsequent measurements with partially overlapping illumination patterns are apparently traded off against more important other benefits identified by the intelligent meta-imager in its learned illuminations. For the moderate-noise and high-noise regimes, we observe that the average overlap *increases* with decreasing $M$, all the way up to 84 % for $M = 2$ and $\rho = 10$. Hence, the intelligent meta-imager deliberately increases the overlap of its illumination patterns as the amount of information that it can extract from the scene becomes more and more limited. This choice can be understood intuitively because subsequent measurements with high overlap are essentially a kind of signal averaging, a well-known human strategy to limit the corruption of measured data through strong noise. However, while signal averaging would correspond to 100 % overlap, the intelligent meta-imager gradually adapts the level of overlap to the noise level and the number of allowed measurements, hence finding an optimized trade-off between some sort of signal averaging and the acquisition of non-redundant information, going well beyond mere signal averaging.

Next, we consider in the third row of Figure 3 the average over the ten realizations of the mean illumination pattern intensity $I$. Recall that the wave energy injected into the TX DMA is always the same throughout this work. For random DMA configurations, the average value of $I$ is again by construction independent of $M$ and $\rho$, and we normalize it to unity for convenience. For learned DMA configurations, even in the low-noise regime ($\rho = 0.1$), as $M$ is decreased from 150, the average intensity increases to almost twice that obtained with random DMA configurations for moderately low values of $M$. Surprisingly, for very small values of $M$ the average intensity drops again, presumably because the limited available control over the illumination pattern is allocated to optimizing the pattern as opposed to the signal strength in this regime. For the moderate-noise regime ($\rho = 1$), the average intensity for learned DMA configurations increases from twice to four times that obtained with random DMA configurations as $M$ is reduced from 150 to 1. In the high-noise regime ($\rho = 10$), the averaged intensity with learned DMA configurations is four times that obtained with random DMA configurations for all considered values of $M$. Again, this choice of the intelligent meta-imager to deliberately increase the intensity is intuitively understandable: as the amount of information that can be extracted from the scene gets more limited due to more noise and/or lower $M$, boosting the signal strength helps to overcome some noise corruption and extract more information per measurement. Human operators routinely choose to focus illumination patterns on regions of interest (ROI) in order to probe them with improved SNR. For instance, in Refs. (*26*, *27*), in a first step a programmable meta-imager illuminates a scene with random patterns in order to identify the ROI; in a second step, waves are then focused on this ROI. In Refs. (*26*,

*27*), the choice to focus in the second step was explicitly imposed by a human operator whereas the intensity increases we observe in Figure 3 occur spontaneously without any human instructions.

How does the intelligent meta-imager increase $I$ by a factor of four on average if the wave energy injected into the TX DMA is always the same? An important aspect are certainly constructive interferences that can be tailored to occur in the scene. However, a second aspect relates to the number of radiating meta-atoms. Clearly, with more radiating elements, on average more energy leaks out of the DMA and toward the scene. While the average ON ratio of the meta-atoms is obviously 50% for random illuminations, it gradually increases up to 68 % as $M$ gets smaller and the noise stronger – see the fourth row in Figure 3. Nonetheless, given that the ON ratio only increases from 50 % to 68 %, the four-fold increase in $I$ must mainly be attributed to tailored constructive interferences.

We have hence discovered two "macroscopic" trends in the learned illumination patterns that can be understood intuitively: as the amount of information that can be extracted gets limited by latency and noise, the intelligent meta-imager tends to increase overlap and intensity of its illumination patterns. While we can intuitively comprehend these choices, these results arise spontaneously in the optimization process without any explicit constraints from us regarding these mechanisms. This observation is reminiscent of other optimization problems in wave engineering where intuitively understandable designs emerged spontaneously; for instance, the inverse design of structures aimed at backscatter-protected transport yielded topological insulators (*28*). Moreover, the insights into the learned illumination patterns give us confidence that average global ("macroscopic") features can be understood intuitively, in contrast to "microscopic" details or realization-specific global features. With respect to the latter, it is interesting to note that while the average accuracy fluctuates little across the ten realizations, the fluctuations in overlap are notable and the fluctuations in intensity are very large. This is also evident upon visual inspection of the provided example illumination-pattern magnitudes in Figure 3. Overall, these findings hint at analogies with statistical physics and thermodynamics, which are indeed known to be intimately connected to machine-learning concepts (*29*).

In Figure 4, we plot the same quantities as in Figure 3 but for a fixed $M = 3$ and sweeping the noise level $\rho$ from $10^{-2}$ to $10^2$. It is apparent that for noise levels below $10^{-1}$, the results become independent of $\rho$ because the noise becomes negligible. For noise levels above 10, we see that the previously discussed trends in terms of overlap and intensity gradually disappear again. This can be attributed to excessively strong noise such that the system cannot successfully operate anymore.

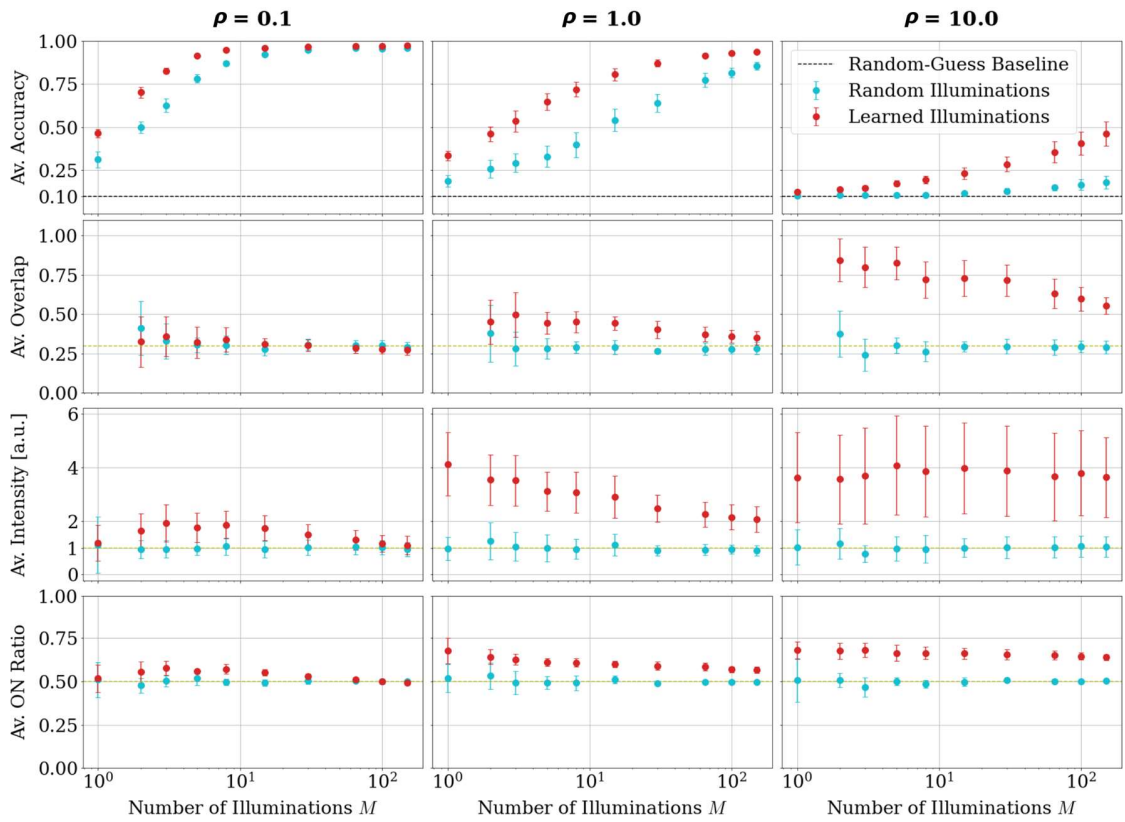

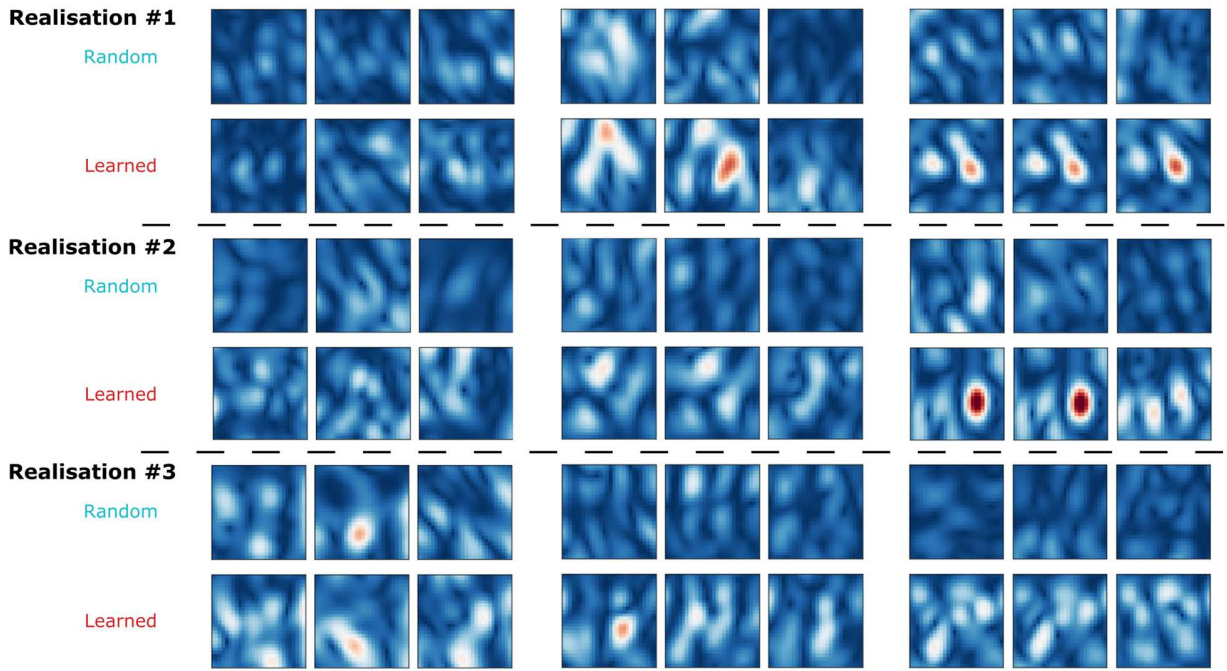

**Figure 3: Intelligent Meta-Imaging Adaptation to Signal-*Independent* Additive Gaussian Noise.** For three representative noise levels ($\rho = 0.1, 1, 10$), the dependence of achieved accuracy (top row), average illumination pattern overlap (second row), average illumination pattern intensity (third row), and average ON ratio of the meta-atoms (fourth row) is plotted as a function of the number of allowed measurements *M*. We contrast our task-specific end-to-end optimized approach ("Learned Illuminations", red) with the conventional compressive-sensing approach of using random DMA configurations ("Random Illuminations",

blue). Each data point represents the average over 10 realizations, each with randomly chosen meta-atom locations. The error bars indicate the standard deviation across these 10 realizations. In addition, the illumination patterns of the first three realizations for $M = 3$ are shown. The colorscale is the same for all displayed illumination patterns.

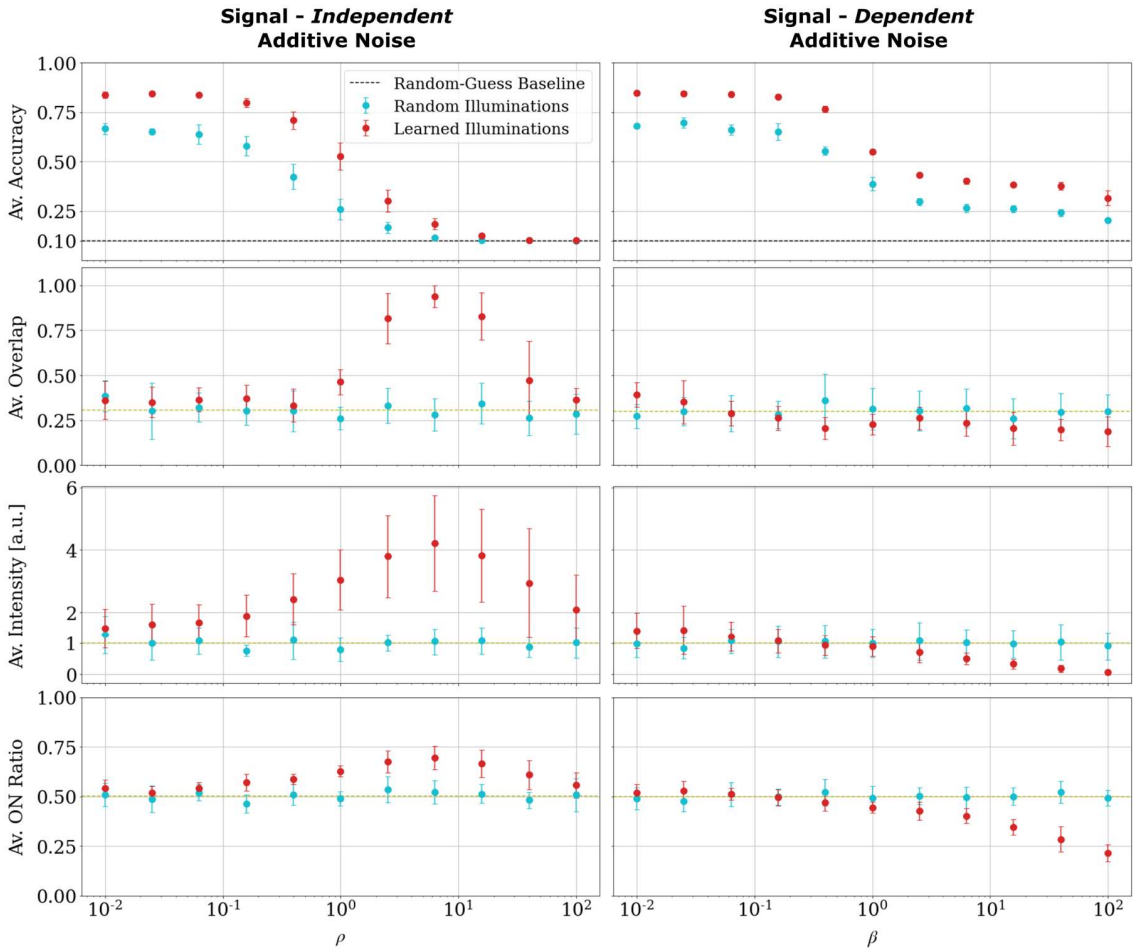

**Figure 4: Intelligent Meta-Imaging Adaptation to Signal-*Independent* vs. Signal-*Dependent* Additive Gaussian Noise.** For $M = 3$, the dependence of achieved accuracy (top row), average illumination pattern overlap (second row), average illumination pattern intensity (third row), and average ON ratio of the meta-atoms (fourth row) is plotted as a function of the noise strength for the two noise types. Each data point represents the average over 20 realizations, each with randomly chosen meta-atom locations. The error bars indicate the standard deviation across these 20 realizations.

**Signal-Dependent Additive Noise.** Having studied signal-*independent* additive noise in the previous section, we now consider a type of noise that is signal-*dependent* additive noise. The purpose of this section is to demonstrate the generality of the noise-adaptiveness of our intelligent programmable meta-imager, and to investigate whether the above-discussed trends for signal-independent noise also hold for other noise types. The signal-dependent noise model we consider draws $\text{Re}(n_i)$ and $\text{Im}(n_i)$ from zero-mean Gaussian distributions with standard deviations $\beta|\text{Re}(m_i)|$ and $\beta|\text{Im}(m_i)|$, respectively. The noise distribution is hence parametrized by $\beta$. In this section, we fix $M = 3$ and sweep $\beta$ from $10^{-2}$ to $10^2$. We do not explicitly claim that this noise model is relevant to a specific microwave experiment, but we do note that this type of noise model may arise, for instance, in certain natural processes (*30*).

The trends observed with this signal-*dependent* noise type in the right column of Figure 4 are qualitatively clearly different from those previously found for signal-*independent* noise in the left column of the same figure. Our intelligent meta-imager hence adapts not only to the noise strength but also to the noise type. With the utilized signal-*dependent* noise model and for $M = 3$, the absolute performance gain with learned illuminations is roughly constant for all considered values of $\beta$. For $\beta \to 0$, this noise model approaches the noiseless regime and hence yields results comparable to the signal-independent noise model from the previous section with $\rho \to 0$. The average overlap slightly decreases as $\beta$ is increased. More notable is that the average intensity approaches zero as $\beta$ is increased. This trend is again intuitively understandable. For high values of $\beta$, strong signals are severely punished with very strong noise such that it appears advantageous to limit the signal intensity. This is achieved through a notable reduction of the ON ratio to below 25 %, but certainly also through tailored destructive interferences in the scene: For $\beta = 10^2$, a quarter of the meta-atoms leaks energy to the scene but the average intensity in the scene is close to zero. In contrast, for $\beta = 10^{-2}$, there is no strong-signal penalty and a mildly enhanced intensity with respect to the use of random DMA configurations is seen in Figure 4.

**Performance Outside the Trained Noise Regime.** Finally, we explore how the recognition performance changes if the intelligent meta-imager is forced to operate outside the trained noise regime. The results displayed in Figure 5 reveal that in the case of signal-*independent* additive Gaussian noise, operating at a noise level that is stronger than in the trained noise regime yields deteriorated performance, as expected. Operating at a lower noise level than during training does not significantly impact the performance. The use of learned as opposed to random illuminations is always advantageous. In the very-strong-noise regime ($\rho > 10$), the average accuracy is close to the random-guess baseline.

A qualitatively different behavior is seen in the case of signal-*dependent* additive Gaussian noise. Here, for low noise levels ($\beta < 1$), we also observe that testing at a noise level lower (higher) than the training noise level does not (does) deteriorate the average accuracy. However, for very strong noise levels ($\beta > 10$), the accuracy is above the random-guess baseline, in contrast to the signal-independent additive noise. Moreover, we observe that in the strong-noise regime the accuracy rapidly falls off as the noise level is detuned in either direction, and detuning toward lower noise levels leads to a faster deterioration of performance. These observations can be understood from the fact that the system can extract some information from the noise because the noise is signal-dependent and hence bears some correlation with the signal. For instance, a high measured value is unlikely to arise from a low signal in the considered signal-dependent noise model. Yet, if the system has carefully adjusted to extract some information from the noise, it is more vulnerable to noise detuning.

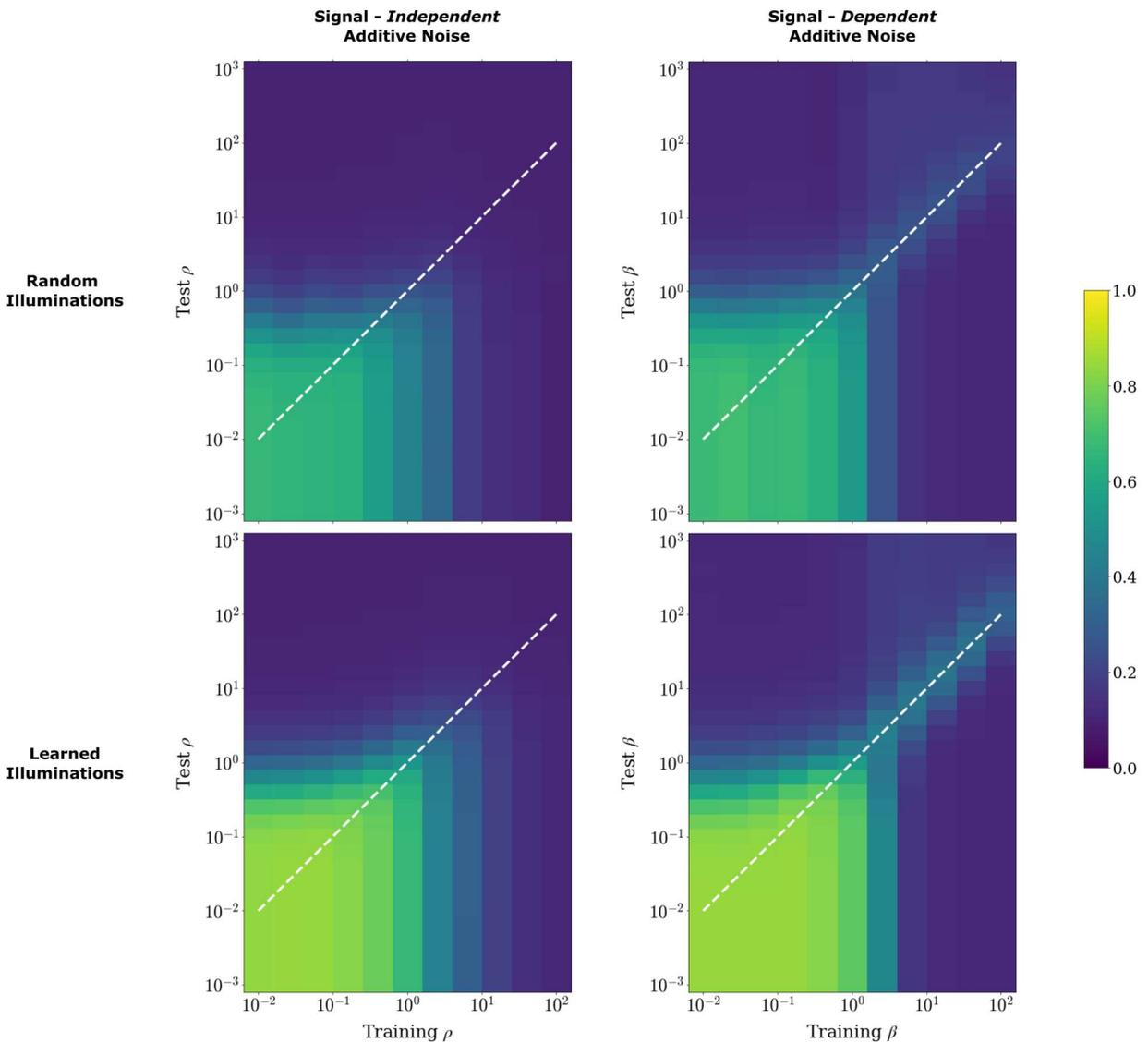

**Figure 5: Performance of Noise-Adaptive Intelligent Meta-Imaging Outside the Trained Noise Regime.** The intelligent meta-imager is adapted to the trained noise regime (horizontal axis) and then tested across a wide range of noise levels beyond the one from the training. For $M = 3$, we show the average accuracy over 20 realizations for random (top row) vs. learned (bottom row) illuminations and the cases of

signal-*independent* (left column) vs. signal-*dependent* (right column) noise. The white dashed lines indicate the training noise regime.

## Discussion

So far, we tested the performance of a programmable meta-imager that generates a sequence of scene illuminations that are specific to a chosen task and a chosen type and level of noise. The transition toward a system that self-adaptively detects the type and level of noise and updates accordingly its utilized sequence of DMA configurations without additional human input is straightforward. First, before runtime, one establishes a codebook of sequences of DMA configurations that are optimized for the various types and levels of noise that one expects to possibly arise during runtime. For instance, in the common case of signal-independent additive complex Gaussian noise, this will involve repeating the end-to-end task-specific optimization for the range of different noise levels that may be encountered. During runtime, the current type and level of noise is simply determined from repeated measurements of the same scene with the same illumination pattern, to then choose accordingly a suitable sequence of DMA configurations from the codebook. This sensing of the noise level should be repeated in regular intervals that correspond to the frequency with which the noise can change in the working environment. Given achievable refresh rates of the DMAs of at least a few tens of kHz, it is possible to perform the noise sensing while a typical scene is effectively static. Therefore, our intelligent programmable meta-imager is not only noise-specific but also straightforwardly noise-adaptive.

Throughout this work, we have focused on different types of detector noise that arise during the detection of a signal. In principle, noise can also arise prior to the signal detection, namely during the signal generation or the signal propagation. Signal-generation noise could arise at the TX DMA and would be multiplexed across the scene illumination patterns, an effect known as noise folding (*31*). Signal-propagation noise would occur if a given instance of the scene reflectivity was not perfectly static but to some extent fluctuating, a phenomenon that may arise, for instance, for operation in dynamically evolving indoor environments as opposed to free space (*32*). Both signal-generation noise and signal-propagation noise cannot trivially be mapped into detector noise, but by accounting for them in the information-flow pipeline, our end-to-end optimization technique would adapt to those types of noise as well.

To summarize, we have presented a noise-adaptive and task-specific intelligent programmable meta-imager, considering a prototypical object-recognition task with microwave DMA hardware.

The amount of information that can be extracted from the scene is limited both through latency constraints and noise in such multi-shot single-detector schemes which are typical for microwave-domain meta-imaging applications. Remarkable performance improvements of our intelligent meta-imager over conventional compressive meta-imagers arise under these conditions. We have demonstrated these performance gains for a signal-independent and a signal-dependent additive noise type. Moreover, we have discovered that "macroscopic" features of the learned illumination patterns, namely their overlaps and intensities, are on average intuitively understandable. We faithfully expect that our results can be transposed to information-extraction problems based on other wave phenomena (e.g., optics, acoustics, elastics, quantum mechanics) and/or with other types of in-situ programmable measurement hardware.